Evaluating the paleomagnetic potential of single zircon crystals using the Bishop Tuff

Roger R. Fu[1,2,3], Benjamin P. Weiss[1], Eduardo A. Lima[1], Pauli Kehayias[4,5], Jefferson F. D. F. Araujo[1], David R. Glenn[4,5], Jeff Gelb[6], Joshua F. Einsle[7], Ann M. Bauer[1], Richard J. Harrison[7], Guleed A.H. Ali[2], Ronald L. Walsworth[4,5]

[1]Department of Earth, Atmospheric and Planetary Sciences, Massachusetts Institute of Technology, Cambridge, MA, USA

[2]Lamont-Doherty Earth Observatory, Columbia University, Palisades, NY, USA

[3]Department of Earth and Planetary Sciences, Harvard University, Cambridge, MA, USA

[4]Harvard-Smithsonian Center for Astrophysics, Cambridge, MA, USA

[5]Department of Physics, Harvard University, Cambridge, MA, USA

[6]Carl Zeiss X-ray Microscopy Inc., Pleasanton, CA, USA

[7]Department of Earth Sciences, University of Cambridge, Cambridge, UK

**Abstract:**

Zircon crystals offer a unique combination of suitability for high-precision radiometric dating and high resistance to alteration. Paleomagnetic experiments on ancient zircons may potentially constrain the earliest geodynamo, which holds broad implications for the early Earth interior and atmosphere. However, the ability of zircons to record accurately the geomagnetic field has not been fully demonstrated. Here we conduct thermal and room temperature alternating field (AF) paleointensity experiments on 767.1 thousand year old (ka) zircons from the Bishop Tuff, California. The rapid emplacement of these zircons in a well-characterized magnetic field provides a high-fidelity test of the zircons' intrinsic paleomagnetic recording accuracy. Successful dual heating experiments on nine zircons measured using a superconducting quantum interference device (SQUID) microscope yield a mean paleointensity of 46.2 ± 18.8 µT (1$\sigma$), which agrees closely with high-precision results from Bishop Tuff whole rock (43.0 ± 3.2 µT). High-resolution quantum diamond magnetic mapping, electron microscopy, and X-ray tomography indicate that the bulk of the remanent magnetization in Bishop Tuff zircons is carried by Fe oxides associated with apatite inclusions, which would be susceptible to destruction via metamorphism and aqueous alteration in older zircons. As such, while zircons can reliably record the geomagnetic field, robust zircon-derived paleomagnetic results require careful characterization of the ferromagnetic carrier and demonstration of their occurrence in primary inclusions. We further conclude that a combination of quantum diamond magnetometry and high-resolution imaging can provide detailed, direct characterization of the ferromagnetic mineralogy of geological samples.

**Introduction:**

Due to its resistance to metamorphism and weathering processes, the silicate mineral zircon ($ZrSiO_4$) preserves a unique record of the Earth's ancient past. Geochemical studies of zircons routinely provide important constraints on their crystallization environment [e.g., Watson & Harrison, (2005)]. Moreover, zircons often provide highly accurate radiometric formation ages due to their high initial U to Pb ratio.

Owing in part to these properties, detrital zircon crystals from the Jack Hills of Western Australia are the oldest preserved terrestrial material and have been dated up to 4.37 Ga (Froude et al., 1983; Harrison, 2009; Valley et al., 2014). These zircons provide the only known opportunity for direct experimental paleointensity measurements on the earliest history of the geodynamo. Such direct constraints on the Earth's early magnetic field hold key implications for a broad range of geophysical problems. A delayed onset of the geodynamo may imply the persistence of a hot, molten lower mantle or the late initiation of plate tectonics (Labrosse et al., 2007; Nimmo and Stevenson, 2000; O'Neill and Debaille, 2014). On the other hand, an active dynamo during the earliest Hadean eon may imply an important role for compositionally-driven convection in the core or a magma ocean overturn event (Elkins-Tanton et al., 2005; O'Rourke and Stevenson, 2016). At the same time, the intensity of the geomagnetic field may have exerted strong control on the rate of atmospheric loss (Kulikov et al., 2007).

Previous paleomagnetic experiments on whole rock samples dating from ~3.5 Ga suggest the existence of an active geodynamo at that age of possibly lower than modern strength (Biggin et al., 2011; Hale and Dunlop, 1984; Hale, 1987; McElhinny and Senanayake, 1980; Tarduno et al., 2010; Yoshihara and Hamano, 2004). On the other hand, the abundance of nitrogen atoms in lunar soil has been interpreted to suggest the lack of a geodynamo at ~3.9 Ga (Ozima et al., 2005). Most recently, a paleomagnetic study of Jack Hills zircons proposed that a geodynamo with lower than modern day intensity has existed since ~4.2 Ga (Tarduno et al., 2015). However, a context study of the Jack Hills area by some of us has questioned the likelihood that a primary paleomagnetic record may be retained in Jack Hills zircons, such that the ages of their magnetizations are currently unknown (Weiss et al., 2015).

Despite the potential of zircon paleomagnetism to expand our understanding of the early geodynamo, the ability of single zircon crystals to record accurately ancient magnetic fields has not yet been tested against bulk rock measurements. Zircon is not ferromagnetic and cannot by itself record ambient magnetic fields. Primary Fe oxide inclusions observed in zircon can potentially carry remanent magnetization (Timms et al., 2012). However, the magnetic recording properties of these and other possible ferromagnetic inclusions in zircon remain unknown. Detailed paleomagnetic and rock magnetic characterization of zircon is therefore necessary to establish the validity of any paleomagnetic record retrieved from zircon grains.

A recent rock magnetic characterization of detrital zircons from the Tanzawa tonalitic pluton, Japan, measured the ratios of natural remanent magnetization (NRM) to thermoremanent magnetization (TRM) in 12 zircons and inferred a mean paleointensity of 74 µT, which overestimates paleofields expected for the sample location based on geomagnetic dipole field intensity records (8-31 µT) by a factor of 2 to 9 (Sato et al., 2015). When filtered to include only samples with high TRM acquisition capacities, the mean paleointensity of eight zircons (41 µT) is broadly consistent with actual geomagnetic fields. However, because no physical relationship was established between rock magnetic properties and the accuracy of paleointensities of individual zircons, the filtering criterion used cannot be extended reliably to future paleointensity studies. Furthermore, because the Sato et al. (2015) study only compares mean recorded fields over the span of several million years, it does not resolve whether discrepancies between individual zircon paleointensities and mean field values are due to paleosecular variations or an inability of zircons to record accurately ambient magnetic fields.

A more robust evaluation of zircon's paleomagnetic recording potential therefore requires detailed paleomagnetic experiments on mineralogically well-characterized zircons with simple geologic histories that acquired NRM in a well-constrained magnetic field. At the same time, zircons erupted in ash flows are preferable to deep-sourced plutonic zircons as the latter experienced prolonged cooling histories, which complicates the comparison of recovered paleointensities directly to known geomagnetic field intensities.

The Bishop Tuff is an extensive sequence of rhyolitic ash fall tuffs and ignimbrites in eastern California (Hildreth, 1979; Wilson and Hildreth, 1997). The melts giving rise to the Bishop Tuff are characterized as evolved silicic magma bodies with high $H_2O$ content (Anderson et al., 1989; Bindeman and Valley, 2002; Wallace et al., 1999). On the basis of inclusion assemblages, trace element concentrations, oxygen isotopic compositions, and Ti-in-zircon thermometry, the Jack Hills zircons are likewise considered to have crystallized from felsic magmas with substantial water content (Hopkins et al., 2010; Mojzsis et al., 2001; Watson and Harrison, 2005; Wilde et al., 2001). The composition of Bishop Tuff zircons may therefore be analogous to that of the Jack Hills samples, although their precise crystallization sequences may have differed given the uncertainties in the crystallization setting of the latter [e.g., Darling et al., (2009); Hopkins et al., (2010); Kemp et al., (2010); Nutman, (2006); Rasmussen et al., (2011)]. The emplacement of the Bishop Tuff occurred in a period of less than a few years at 767.1 ka, which postdates the Matuyama-Bruhnes reversal (Crowley et al., 2007; Singer et al., 2005; Snow and Yung, 1988; Wilson and Hildreth, 1997). As such, paleomagnetic records from the deposit were not subject to viscous overprinting in a reversed geomagnetic field or post-depositional metamorphism. Meanwhile, the cooling period of TRM acquisition, likely on the order of tens to a few hundred years (Riehle et al., 1995), was too short to result in internal heterogeneities within the deposit due to paleosecular variations.

A previous paleomagnetic study of the Bishop Tuff showed that densely welded ignimbrite samples contain a primary TRM most likely carried by primary low Ti titanomagnetite (Gee et al., 2010). Thellier-Thellier paleointensity experiments performed on 61 whole rock samples showed that approximately 75% yielded reliable estimates of the paleomagnetic field intensity. The narrow range of paleointensity values derived from samples recovered from diverse locations and stratigraphic heights (43.0±3.2 µT) confirms that cooling through the temperatures of NRM acquisition occurred sufficiently fast to avoid recording paleosecular variations. Because of the mineralogical properties described above, the simple geologic history, and the availability of detailed characterization of the paleofield intensity during emplacement, Bishop Tuff zircons offer a unique opportunity to evaluate the quality of primary paleomagnetic recording in zircon.

In this study, we conducted thermal and AF paleointensity experiments and characterized the ferromagnetic mineralogy of zircons from the Bishop Tuff. We chose zircons taken from sites adjacent to the locations of samples suggested by Gee et al. (2010) to carry primary TRMs. To acquire paleomagnetic data from single zircons, we developed and describe here novel techniques that permit thermal demagnetization and measurement of samples with NRM magnitude as low as $5\times10^{-14}$ Am$^2$. This sensitivity represents a gain of greater than one order of magnitude for samples subject to thermal demagnetization compared to previous techniques (Fu et al., 2014b; Tarduno et al., 2015). Comparison between single zircon and whole rock thermal paleointensities shows that the two data sets agree to within uncertainty. Results from our magnetic and electron microscopy and X-ray tomography suggest that stable magnetization in Bishop Tuff zircons is carried by Fe-oxides associated with apatite inclusions.

**Samples and methods:**

We collected samples from an exposure of the Bishop Tuff from the Owens River Gorge. The sampling location (37.51189°N, 118.57129°W) is approximately 50 m north and 32 m up-section from the base of the ~150 m thick Gorge Section F (GF) studied by Gee et al. (2010). All of our zircons are therefore found within the dense welded ignimbrite unit Ig1Eb (Wilson and Hildreth, 1997), which oxygen isotopic data suggest experienced limited to no post-depositional hydrothermal alteration (Holt and Taylor, 1998). Assuming that the flow unit was deposited isothermally, its mass density provides an emplacement temperature estimate of ~660°C (Gee et al., 2010), although the uncertainty on this temperature estimate is not well-understood. Variations in observed ignimbrite density ~25 m up-section from our sampling locality may imply non-uniform deposition temperatures; however, such heterogeneities likely do not significantly affect the assumption of isothermal deposition in the lower part of the Ig1Eb unit where our samples were taken (Wilson and Hildreth, 2003; Riehle et al., 2010). As such, the emplacement temperature implies that our zircons should have recorded a nearly full TRM, although a weaker component of hematite-hosted remanence possibly acquired via chemical alteration may exist at higher temperature for some samples (Gee et al., 2010). The close proximity of our samples to the sites used in Gee et al. (2010) implies that the samples experienced similar cooling histories and that a comparison of recovered paleointensities would not require corrections for cooling rate (Yu, 2011). Finally, the location of the samples in a steep-sided canyon minimizes the potential of lightning contamination, which is confirmed by our paleointensity results (see below).

We crushed 2 kg of ignimbrite and separated 40 zircons without the use of a magnetic separator. Isolated zircons, which range between 90 and 180 μm in the longest dimension, were handled using plastic and ceramic tools and kept inside the MIT Paleomagnetism Laboratory class ~10,000 clean room (DC field <150 nT). For AF demagnetization experiments, we measured 15 zircons embedded in EPO-TEK 301 epoxy extruded through a 0.25 μm filter (see

Tables 1 and S1 for lists of samples). The surface of this epoxy mount containing the zircons was polished using 1 μm alumina powder. For thermal demagnetization experiments, we embedded 18 zircons in pits drilled into Corning Eagle XG glass slides and secured the zircons using high purity quartz powder (Fig. 1). In both cases, the placement of samples in epoxy or quartz powder insured consistent orientation during the demagnetization process, although the zircons were not oriented in an absolute coordinate system.

Due to the very weak initial NRM moments of zircons ($5\times10^{-14}$ to $6\times10^{-12}$ Am$^2$), we measured all samples using the superconducting quantum interference device (SQUID) microscope in the MIT Paleomagnetism Laboratory (Weiss et al., 2007). For AF demagnetization samples, the polished surface of the epoxy mount was placed in direct contact with the SQUID microscope window, resulting in a sensor-to-sample distance of 180 μm. For thermal demagnetization samples, we placed the SQUID microscope window in contact with the undrilled side of the glass slides, yielding sensor-to-sample separations of ~230 μm. In both sets of experiments, the SQUID microscope produced magnetic field maps with ~200 μm spatial resolution. We computed zircon magnetic moments from these magnetic field maps by forward modeling of a dipole magnetic source and optimization of the model dipole parameters so as to match the experimental data. Specifically, we adjusted the three components of the magnetic dipole moment and the spatial location of the source to minimize residuals between the measured magnetic field map and the synthetic field map calculated from the model [for more details, see Fu et al. (2014a)].

The epoxy and glass mounts, when measured without samples, displayed a small number (1-2 per mount) of contaminating magnetic sources with moment $1\text{-}7\times10^{-14}$ Am$^2$. Although the upper end of this range overlaps with the weakest measured zircon moments, SQUID microscopy showed that the location of these sources do not coincide with that of measured zircons and therefore do not affect our computed zircon magnetizations (Fig. 1). At the locations of the mounted zircons in the magnetic field map, temporal fluctuations in the SQUID microscope sensor resulted in noise of up to 100 pT (Fig. 1). Assuming that the signal from a zircon must exceed this field strength to constitute a robust measurement, a zircon must have in this particular configuration (sensor-to sample separation of 230 μm) a dipole moment of $>1.2\times10^{-14}$ Am$^2$. Because this calculation does not take into account improvements in signal-to-noise ratio from signal processing techniques and the fact that the optimization algorithm can retrieve dipole moments at higher noise levels, the value of $1.2\times10^{-14}$ Am$^2$ represents a conservative estimate of our ability to characterize a zircon signal from a raw SQUID microscope map. This moment is higher than that given in previous SQUID microscope-based studies (Fu et al., 2014b; Weiss et al., 2007) due to the higher (230 μm) sensor-to-sample separation necessitated by the thermal demagnetization procedure.

We performed further rock magnetic characterization of selected zircons after their NRMs were removed via AF demagnetization. We used a JEOL 8200 electron microprobe in the MIT Petrology Laboratory set at an accelerating potential of 15 keV to image and characterize ferromagnetic mineralogy. To complement and guide these analyses, we mapped the location of ferromagnetic sources in the zircons using the quantum diamond microscope (QDM) developed at the Walsworth Laboratory at Harvard University. This instrument, which was first used on geological samples to localize ferromagnetism in dusty olivine-bearing chondrules from the Semarkona meteorite (Fu et al., 2014b), produces maps of the magnetic field above a polished sample with a spatial resolution of 2.4 μm and DC noise floor of ~160 nT. The QDM employs a dense layer of fluorescent quantum magnetic field sensors, nitrogen vacancy color centers, near the surface of a diamond chip on which the sample of interest is placed (Glenn et al., 2015).

Finally, we used the ZEISS Xradia 520 Versa and Ultra XRM-L200 from Xradia, Inc. (now Carl Zeiss X-ray Microscopy) microscopes to obtain tomographic volume reconstructions of a single zircon grain (A7) with 400 nm and 65 nm voxel size, respectively. These voxel sizes give spatial resolutions of 750 nm and 150 nm, respectively, based on measured test patterns using visual inspection and a Modular Transfer Function. We extracted this zircon, which was subjected earlier to AF demagnetization, QDM, and electron microprobe analyses, from the epoxy holder and remounted it onto a ~50 μm needle tip with superglue for X-ray tomography studies. Both X-ray microscopes were used in absorption contrast mode, in which the contrast roughly scales with mass density.

The sub-micrometer Versa XRM was operated at an accelerating voltage of 60 keV and power of 5 W using the full spectrum of the X-ray beam; the nanoscale Ultra XRM was operated at a quasi-monochromatic X-ray energy of 8 keV, above the Fe absorption edge, which produced a large contrast signal for the Fe oxide particles compared to other materials in the volume studied. For the sub-micrometer-scale Versa XRM imaging, 1601 images were collected at a step size of 0.23 degrees for a total acquisition time of 2 hours; for the nanoscale XRM tomographic series, 901 projection radiographs were acquired with an angular resolution of 0.2 degrees and a total scan time of 13 hours. After acquisition, the two series of radiographs were aligned and reconstructed using the ZEISS XMReconstructor software. Visualization and analysis of the resulting three-dimensional (3D) volumes were performed using ORS Visual Si Advanced and Zeiss XM3DViewer.

**Results:**

*Paleomagnetic measurements*

We performed thermal demagnetization and paleointensity analysis on 18 zircons whose NRM intensities range between $5.2 \times 10^{-14}$ and $5.7 \times 10^{-12}$ Am$^2$ with a mean of $1.3 \times 10^{-12}$ Am$^2$ and median of $9.1 \times 10^{-13}$ Am$^2$. For ten zircons composing the T2 group (Tables 1-2), we implemented the IZZI protocol where samples are heated in a 0 µT and a 50 µT field for each temperature step in alternating order (Yu et al., 2004). We heated samples up to 673°C in 25 steps, with 10°C steps between 520°C and 583°C. Partial TRM (pTRM) checks were performed at seven temperatures between 269°C and 583°C. For eight zircons composing the T1 group, the full thermal demagnetization sequence, which consisted of 20 steps up to 680°C, was completed before pTRM acquisition experiments in order to acquire demagnetization sequences of the NRM with the least possible amount of sample alteration. The data for these eight zircons therefore did not include pTRM checks. All heating was conducted in air for 30 minutes at the peak temperature, although the isolation of zircons by a ~1 mm layer of packed quartz powder may have limited chemical exchange with the oven environment.

We used directional demagnetization data (i.e., orthogonal projection diagrams; Fig. 2) to identify the highest fidelity range of demagnetization steps for each zircon for paleointensity calculations. Seventeen out of eighteen zircons subjected to thermal demagnetization carry a resolvable component of magnetization with maximum angle of deviation (MAD) less than 25° [Table 1; (Kirschvink, 1980)], although directionally coherent magnetization in zircon T2-10 was blocked below only 150°C. Because the Bishop Tuff was deposited after the Brunhes-Matuyama reversal, we do not observe overprints of reverse polarity in our demagnetization sequences. However, Gee et al., (2010) observed a high temperature (>580°C) component of magnetization possibly due to chemical alteration and showing anomalous paleointensities. At the same time, pre-depositional remanence is potentially retained in zircons at the highest temperatures. Due to the unknown paleointensity recording efficiency of chemical remanence and possible pre-deposition tumbling of the zircons, we exclude components of magnetization that are blocked exclusively between 550 and >600˚C. Such high temperature components of remanence are observed in six zircons (Fig. 2; Table 1). The differences in directions of these components and those unblocked at lower temperatures suggest that the former represent pre-depositional magnetizations or post-depositional chemical remanent magnetizations carried by hematite that formed at a later time.

We compare the MAD and deviation angle (DANG) of observed components of magnetization to test whether they trend to the origin [Table 1; (Tauxe and Staudigel, 2004)]. For nine out of seventeen zircons with resolvable components, the value of DANG for the magnetization component chosen for paleointensity analysis is less than or approximately equal to MAD, suggesting that these components are origin-trending. Of the remaining eight zircons, four (T1-1, T1-5, T1-7, and T2-8) carry higher temperature magnetizations (see above). The lower temperature components analyzed for paleointensity in these zircons are therefore not expected to be origin-trending. For a final group of four zircons (T1-2, T1-8, T2-2, and T2-4), noisy demagnetization sequences at higher temperatures may confound the isolation of a higher-temperature component. In any case, only one of these last four zircons (T1-8) passes the quality criteria adopted for computing the mean paleointensity (see below). As such, eight out of nine zircons ultimately selected to determine the mean paleointensity have their paleointensities determined from either origin-trending magnetizations or carry components at higher temperatures.

Eleven out of seventeen zircons with resolvable components have a maximum unblocking temperature above 580°C, indicating the presence of hematite formed during post-emplacement or syn-emplacement oxidation [Fig. 3; (Gee et al., 2010)]. Of the remaining zircons, four have maximum unblocking temperatures between 500°C and 583°C, suggesting magnetite as the dominant ferromagnetic phase. One zircon (T2-4) has no resolvable magnetization components above 444°C which may be due to the lack of ferromagnetic grains at higher unblocking temperatures or chemical alteration during laboratory heating.

We computed paleointensities for the sixteen zircons with identified high temperature magnetization components using a linear least-squares fit (Fig. 4). The zircon T2-10 was excluded from this analysis due to its very low unblocking temperature. We then evaluated the fidelity of each paleointensity experiment using the normalized error of the best-fit slope [$\sigma/|b|$; (Selkin & Tauxe 2000)], the fraction of NRM intensity represented in the fitted data range ($f$), the gap factor representing the evenness of NRM loss at each demagnetization step ($g$), and the quality index [$q$; (Coe et al., 1978)]. Thirteen out of sixteen zircons yielded a value for $f$ greater than 0.5 and a least-squares fit range that includes four or more data points, suggesting that the paleointensities are based on a sufficiently large fraction of the sample magnetizations [Table 2; (Coe et al., 1978)]. On the other hand, zircons displayed a wide range of $\sigma/|b|$ between 0.04 and 0.63. Although paleointensity studies of well-behaved whole rock samples typically adopt a value in the 0.05-0.10 range as the maximum acceptable $\sigma/|b|$ (Gee et al., 2010; Tauxe et al., 2013), the very weak magnetizations of our zircons led to significantly greater scatter compared to high quality whole rock samples. We therefore included all zircons with $\sigma/|b| < 0.25$ in our mean paleointensity calculation. After the application of this quality filter the number of zircons included in the final mean paleointensity calculation was reduced from sixteen to ten (Table 2).

Finally, we excluded the zircon T2-9 as an outlier due to its large (4.3$\sigma$) deviation from the mean paleointensity of the remaining zircons. Unlike all other samples, this zircon carries a single, unidirectional component of magnetization blocked between room temperature and 640˚C, which may suggest remagnetization post-dating the formation of hematite. Exposure of the zircon to a strong magnetic field is simultaneously consistent with its single, high temperature component of magnetization and anomalously high paleointensity. Remagnetization due to lightning-induced magnetic fields is unlikely since all zircons samples were extracted from a ~10 cm block sample. As such, exposure of this zircon to a laboratory magnetic field, possibly during zircon separation, is the most plausible explanation for its strong paleointensity. For the final group of nine zircons that pass all reliability criteria above, the mean value for the quality index is 4.38 while the mean MAD is 15.1˚ with a maximum of 20.1˚. These zircons yield a raw mean paleointensity of 46.2 ± 18.8 µT (1$\sigma$) and a weighted mean paleointensity of 40.0 ± 1.3 µT using $1/\sigma^2$ weighting (Taylor, 1997). Both of these values are in agreement with the 43.0 ± 3.2 µT result from higher fidelity bulk Bishop Tuff material [Fig. 5; (Gee et al., 2010)].

Although we conducted repeated pTRM acquisition experiments on ten zircons to check for chemical alteration, the inherently large-scatter demagnetization sequences of the weak zircon samples complicate the interpretation of these results (Table 3). In addition to measurement scatter, the smaller number of ferromagnetic grains contained in each zircon sample in comparison to a standard ~2.5 cm paleomagnetic core increases the observed scatter during demagnetization and may result in ~10% statistical uncertainty in paleointensities derived from zircons with our observed range of NRM moments (Berndt et al., 2016). A standard criterion to evaluate the degree of alteration is the difference ratio (DRAT), which normalizes the difference in the original and repeated pTRMs acquired at a given temperature ($\Delta pTRM_{max}$) by the length of the line segment fitted across the temperature range used to compute paleointensity (Selkin and Tauxe, 2000). For well-behaved, whole rock samples, maximum DRAT values greater than 0.05 or 0.10 have typically been adopted as a sign of sample alteration. However, the original DRAT parameter was defined for strongly magnetized samples where the scatter of each measurement is small compared to the change in magnetization between temperature steps. In contrast, the scatter observed in the Arai diagrams for individual zircons is often comparable to the change between temperature steps (Fig. 4). Because the observed value of $\Delta pTRM_{max}$ contains contributions from both intrinsic scatter and true change in sample pTRM acquisition due to alteration, the canonical DRAT parameter may significantly overestimate the effect of sample alteration in zircons and other weakly-magnetized samples. As a result, the maximum DRAT parameter calculated for our zircons exceeds 0.10 for all samples except T2-9.

As an alternative, we computed the quantity $\sigma_{pTRM}$, which we define as the standard deviation of the difference between the observed pTRM gained and the best-fit value derived from the Arai diagram (i.e., the *x*-coordinate difference between data points on the Arai diagram and the best-fit line). Under the assumption that a linear trend exists in the Arai diagram, $\sigma_{pTRM}$ estimates the amount of variation in pTRM acquisition that is due to measurement noise. We then computed the ratio $\Delta pTRM_{max}/\sigma_{pTRM}$ to determine whether the observed value of $\Delta pTRM_{max}$ may be attributed to measurement scatter. A value of $\Delta pTRM_{max}/\sigma_{pTRM}$ greater than 2 indicates that the observed change in repeated pTRM acquisitions is unlikely (probability $P<0.05$) to be due to noise alone and suggests the onset of sample alteration. By this criterion, three out of ten zircons experienced significant alteration between 487°C and 673°C (Table 2). When considering only pTRM checks within the temperature range used to compute paleointensities, we find that all zircons pass this quality criterion. However, as described above, our simple method of calculating $\sigma_{pTRM}$ may overestimate pTRM scatter as it does not account for nonlinearity of data points in the Arai diagram. Future studies of similarly weak samples may benefit from performing two or more repeated pTRM acquisition experiments after the same higher temperature step to provide an estimate of the inherent noise in pTRM measurements.

For AF-treated zircons, the close (1.5 mm) separation of the 15 zircons on the epoxy holder resulted in contaminated field signals for some zircons mounted adjacent to much stronger samples. As such, reliable magnetic moment inversions were recovered from 12 of the 15 zircons. For these, initial NRM intensities range between $1.9\times10^{-13}$ and $6.2\times10^{-12}$ Am$^2$ with a mean of $2.3\times10^{-12}$ Am$^2$. AF application yielded significantly noisier demagnetization sequences compared to thermal demagnetization (Fig. 2F). We applied AF in 5 mT steps up to 20 mT and in 10 mT steps between 20 and 140 mT. All AF demagnetization sequences showed one identifiable component of magnetization that is fully removed by AF application of between 70 and 130 mT. To derive paleointensities using the IRM normalization method, we imparted a near-saturation IRM of 0.4 T and conducted AF demagnetization using the same steps as described above up to 90 mT. We then computed the least-squares best-fit line to a diagram of NRM remaining and IRM removed to find the ratio of NRM to IRM in the 0 to 90 mT range (Fig. 4F; Table S1).

The mean NRM to near-saturation IRM ratio (*NRM/IRM*) for 12 zircons is 0.15, which may be converted to a paleointensity ($B_{paleo}$) estimate using the relationship:

$$B_{paleo} = f_{IRM}(NRM/IRM)$$

where $f_{IRM}$ is an empirically calibrated factor. Adopting $f_{IRM}$ = 3000 µT based on experiments on magnetite and titanomagnetite with a wide range of domain states (Gattacceca and Rochette, 2004; Kletetschka et al., 2003), we derive a paleointensity of 437 ± 314 µT, which is larger than the expected value by a factor of 10 (Gee et al., 2010). This result implies that Bishop Tuff zircons have anomalously low values of $f_{IRM} \cong 300$ µT, which corresponds to $NRM/IRM \cong 0.17$ for a TRM acquired in an ambient field of 50 µT.

As such, the IRM normalization method failed to recover accurate paleointensities from individual zircons, in contrast to dual heating techniques. The values of $NRM/IRM$ from our analysis are broadly consistent with the range (0.01 to 1) observed in Tanzawa pluton zircons (Sato et al., 2015). Similarly high values of $NRM/IRM$ have been observed in natural single-domain magnetite particles embedded in anorthosite (Kletetschka et al., 2006). At the same time, although observed $NRM/IRM$ values in synthetic magnetite-bearing samples are lower by at least 40% compared to our results, grain sizes in the 100 nm to 1 µm range represent the closest approximation (Yu, 2006). These small inferred grain sizes are consistent with our results from electron microscopy (see below).

*Microscopy of Bishop Tuff zircons*

After the completion of AF demagnetization and IRM acquisition experiments described above, we applied magnetic and electron microscopy techniques to zircons A1-A15 to characterize their ferromagnetic mineralogy. We first use the high-resolution magnetic field imaging capability of the QDM to localize the sources of ferromagnetism (Fig. 6). By imparting an out-of-plane IRM of 0.4 T and placing the nitrogen-vacancy-bearing diamond chip against the polished surface of the epoxy mount, we mapped the magnetic field at a sensor-to-sample distance of ~1 μm from the zircon surface. Locations with strong out-of-plane magnetic fields indicate the presence of ferromagnetic material at the surface or sub-surface. The highly non-uniform spatial distribution of magnetic field sources suggests that ferromagnetic phases are concentrated in inclusions inside the zircon or in uniquely Fe-rich sub-volumes of the zircon grain itself. These observed distributions of magnetic fields are not consistent with magnetite grains uniformly dispersed throughout the zircon [e.g., Timms et al. (2012)], which would result in a spatially broad magnetic field across the full exposed surface of the zircon.

We can place an order-of-magnitude upper bound on the contribution to the total paleointensity from hypothetical volumetrically dispersed ferromagnetic sources (which we refer to as the zircon's "bulk magnetization"). To do this, we selected a zircon (A15) with few localized sources and performed magnetic moment inversions on four localized sources. We then subtracted the dipolar magnetic field associated with each source from the QDM map (Fig. S1A-B; Supplementary Text). The residual magnetic fields after subtracting contributions from these localized sources should be due to the bulk magnetization, if any. To compute the moment of any bulk magnetization, we approximated the zircon as a uniformly magnetized sphere with a 40 μm radius. Because the external magnetic field of a uniformly magnetized sphere is equivalent to that of a dipolar source situated at its center (Dunlop and Ozdemir, 1997), we fitted a dipolar magnetic source to the residual magnetic field map assuming that the dipole is situated at the center of the zircon at 40 μm depth (Fig. S1C). This fit yields a moment of $8.7 \times 10^{-14}$ Am$^2$ for the near-saturation IRM associated with the hypothetical zircon bulk magnetization, which is weaker than the $2.0 \times 10^{-13}$ Am$^2$ moment from the strongest localized source. Because some of this signal may be due to subsurface, localized ferromagnetic sources, we regard $8.7 \times 10^{-14}$ Am$^2$ as an upper bound to the bulk magnetization contribution to the IRM. Furthermore, since the expected NRM moments are weaker by a factor of ~7 than the IRM (see above), the maximum NRM moment contributed by bulk magnetization is $1.2\times10^{-14}$ Am$^2$. We therefore conclude that volumetrically dispersed sources in the whole zircon grain contribute at most ~1% of the total moment of typical zircons with NRM moments of ~$1\times10^{-12}$ Am$^2$ and that the majority of zircon ferromagnetism is concentrated in localized regions of the zircon grain.

We further note that the peak field intensity as measured by the SQUID microscope associated with zircon A15 carrying a 0.4 T IRM is 12 nT. Given that the sensor-to-sample distance of the SQUID microscope is 180 μm and again assuming a dipolar source at 40 μm depth, the expected peak field contribution from the bulk zircon magnetization in the QDM map is 2.0 μT if the zircon's magnetic signal in the SQUID microscope is due to a bulk magnetization. The absence of such a signal from the QDM map therefore confirms that any bulk magnetization cannot constitute a significant fraction of the observed remanence.

To characterize the localized sources of remanent magnetization, we overlaid the QDM magnetic field maps on backscatter electron (BSE) images of the zircons with a registration accuracy of better than 10 μm. The comparison of QDM magnetic field and BSE images shows that the carriers of remanent magnetization are strongly associated with inclusions of apatite (Fig. 6). Wavelength dispersion spectroscopy (WDS) element mapping of Fe distribution confirms that Fe-bearing material occurs within and around apatite inclusions, which may correspond to the observed magnetite and hematite remanence carrier populations, respectively.

We performed 3D X-ray tomography on zircon A7 to compare the spatial distributions of observed magnetic fields with the positions of accessory phases such as apatite within the zircon. We chose zircon A7 in part due to the presence of a strong magnetic signal detected by the QDM that does not correspond to any surface inclusions imaged by the electron microprobe (Fig. 6A), which suggests the presence of interior ferromagnetic phases. X-ray tomography of this sample indeed shows a group of apatite inclusions that correspond closely to the magnetic field source (Fig. 7A-C; Supplementary Data).

Subsequent higher magnification scans using the nanoscale XRM of a 50 μm region (green box in Fig. 7A,B) reveal the presence of high X-ray absorption grains (candidate Fe-oxides) both within the apatite inclusion and at the boundary between the apatite and its zircon host (Fig. 7D). To summarize the results of zircon microscopy, a combination of surface BSE and WDS mapping and 3D X-ray tomography shows that all magnetic signals observed using the QDM correspond to exposed or interior apatite inclusions.

**Discussion:**

Our paleomagnetic measurement of 30 Bishop Tuff zircons showed that all specimens carry resolvable remanent magnetization with NRM intensities between $5\times10^{-14}$ and $6\times10^{-12}$ Am$^2$. Thermal demagnetization and dual-heating paleointensity experiments on 18 zircons indicate that, once filtered using several reliability criteria, approximately 50% of individual zircons are capable of accurately recording a primary geomagnetic field with $1\sigma$ uncertainties of 18.8 μT. The agreement between our single zircon paleointensities and those inferred from bulk samples demonstrates the lack of contamination during our analysis procedures. The apparent ubiquity of ferromagnetic inclusions in zircon capable of carrying paleomagnetic information suggests that, given adequate preservation, single zircon crystals are a potentially robust source of constraints on ancient magnetic fields.

Our microscopy of Bishop Tuff zircons shows that ferromagnetic minerals are exclusively associated with apatite, which is the most common inclusion found in zircon crystals (Jennings et al., 2011). Embedded, euhedral grains of apatite similar to those observed in our Bishop Tuff zircons have been observed in zircons extracted from a wide range of whole rock compositions (Corfu et al., 2003). Furthermore, titanomagnetites in silica-rich magmas may crystallize at temperatures as high as ~1100°C and continue to form over a wide range of temperatures (Ghiorso and Sack, 1991). In comparison, apatite saturates over a lower range of temperatures between 700 and 900°C for similar melts (Harrison and Watson, 1984), permitting the inclusion of early-formed titanomagnetites in apatite. As expected for the host phase, zircon in the Bishop Tuff continued to crystallize at temperatures as low as ~720°C (Reid et al., 2011). As such, the localization of such ferromagnetic grains in included apatites as observed in the Bishop Tuff zircons may be common in silicic, zircon-bearing rocks, leading to a broadly applicable mineralogical criterion for the preservation of primary remanent magnetization in ancient zircons. In this case, the potential of a given population of zircons to retain a primary paleomagnetic record would depend strongly on the preservation of primary apatite inclusions.

Because apatite is susceptible to dissolution and replacement by aqueous fluids (Nutman et al., 2014), ancient zircons that have undergone metasomatism may not retain a primary paleomagnetic record. In the case of the Jack Hills zircons, comparison to modern zircons from similar crystallization environments suggests that approximately 90% of apatite inclusions have been removed during secondary metasomatic events (Rasmussen et al., 2011). However, although only ~5% of Jack Hills zircons are observed with apatite inclusions, all such inclusions are isolated from cracks and thereby show no evidence for a secondary origin (Bell et al., 2015; Yamamoto et al., 2013). At the same time, electron microscopy of ~3.3 Ga Jack Hills zircons has revealed Fe-oxide, in the form of hematite, precipitated during secondary alteration in highly damaged zones (Utsunomiya et al., 2007).

Synthesizing these observations with our results, primary remanent magnetization is potentially retained in select sub-volumes of Jack Hills zircons with unaltered apatite inclusions. However, given the rarity of apatite inclusions and the known presence of secondary iron oxides, careful attention, including high-resolution magnetic microscopy, must be devoted to establishing the locations and the primary nature of ferromagnetic inclusions before any primary record of the Earth's early magnetic field can be accepted as robust.

**Conclusion:**

Using the SQUID microscope and a new measurement protocol with a noise floor of $\sim 1\times 10^{-14}$ $Am^2$ in magnetic moment, we demonstrate that all 30 measured zircons extracted from the Bishop Tuff contain ferromagnetic inclusions that retain robustly recoverable remanent magnetization. Thermal demagnetization and paleointensity experiments on 18 zircons yielded nine specimens that pass a set of reliability criteria based on the fraction of NRM used to derive paleointensity, the observed scatter in the best-fit paleointensity, and the degree of alteration during laboratory heating. The raw and weighted mean paleointensities derived from these nine zircons are 46.2 ± 18.8 µT and 40.0 ± 1.3 µT, respectively, which are in agreement with the result of 43.0 ± 3.2 µT obtained from high-fidelity bulk samples of the Bishop Tuff (Gee et al., 2010). AF demagnetization and IRM normalization paleointensity experiments on 11 zircons showed a higher degree of scatter than thermally demagnetized samples and suggest an inherently high ratio of NRM to near-saturation IRM corresponding to 0.17 in an ambient field of 50 µT. Magnetic microscopy using the QDM, combined with electron microscopy and X-ray tomography, showed that all detectable sources of ferromagnetism in Bishop Tuff zircons are hosted within or around apatite inclusions while typically <1% may be carried by ferromagnetic grains uniformly dispersed within the zircon itself.

These results suggest that thermal demagnetization and paleointensity experiments on zircons may yield reliable estimates of ancient geomagnetic field intensities. However, the retention of primary paleomagnetic information is contingent on the preservation of zircon inclusions, most importantly of apatite. Paleomagnetic studies of single zircon grains, including the Jack Hills zircons (Tarduno et al., 2015), must therefore establish the primary nature of the ferromagnetic carriers before a robust characterization of ancient magnetic fields is possible. Such a demonstration has not yet been provided by any study. Micrometer-resolution magnetic imaging using the QDM, combined with electron and X-ray imaging technique, can play a critical role in identifying the ferromagnetic remanence carriers in other zircon populations and geological samples.

**Acknowledgments:**

We thank Samuel A. Bowring, Timothy L. Grove, T. Mark Harrison, Dennis V. Kent, Simon Lock, and Joseph G. O'Rourke for insight and discussions that improved the content of this contribution. We also thank Eric Barry and Andrew Gregovich for zircon separation, Nilanjan Chatterjee for electron microprobe assistance, and Chenchen Luo for help with the QDM measurements. Finally, we would like to thank Mr. Simon Doe and Mr. Christopher Bassell for arranging access to the XRM instrumentation at the ANFF SA node. RRF is supported by the Lamont-Doherty Post-Doctoral Fellowship. PK acknowledges support from the IC Postdoctoral Research Fellowship Program. DRG and RLW acknowledge support from the DARPA

QuASAR and NSF EPMD programs. JFE and RJH acknowledge funding under ERC Advance grant 320750- Nanopaleomagnetism. This work was performed (in part) at the South Australian node of the Australian National Fabrication Facility under the National Collaborative Research Infrastructure Strategy to provide nano and microfabrication facilities for Australia's researchers.

**Figures and Tables:**

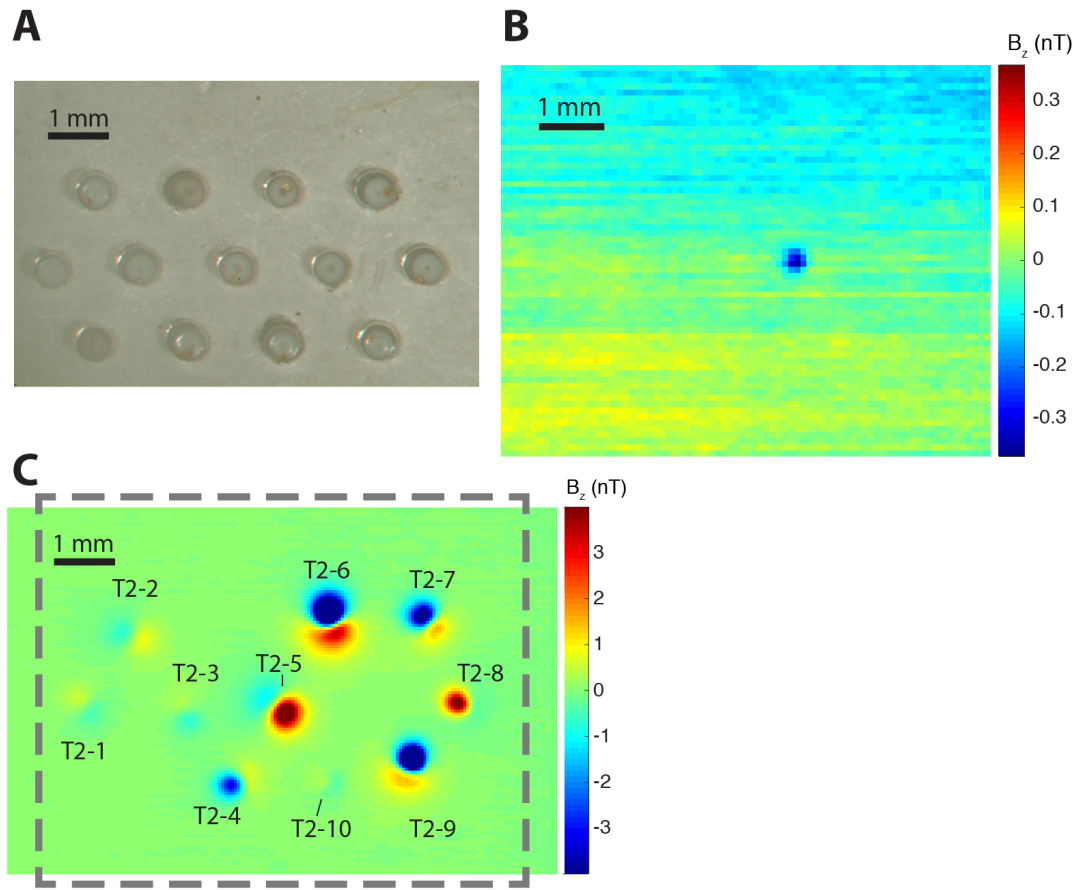

**Figure 1:** (**A**) Optical photograph of the glass sample holder with zircons T2-1 through T2-10 mounted inside pits, (**B**) SQUID microscope map of the same sample holder with no zircons, and (**C**) SQUID microscope map of the sample holder with T2 zircons mounted. Shown is the vertical component of the magnetic field at a height of ~230 μm, with positive (negative) field values representing out-of-plane (into-plane) field directions. The magnetic field signature associated with each zircon is labeled (notice that due to variability in holder fabrication, not every pit was used to house a zircon). Gray dashed box in panel C indicates the area of the map in panel B. Note the different field strength (color) scales between panels B and C. All three

panels have the same scale bar. Complete raw data, processed products, and additional documentation of procedures are available in the Supplementary Data.

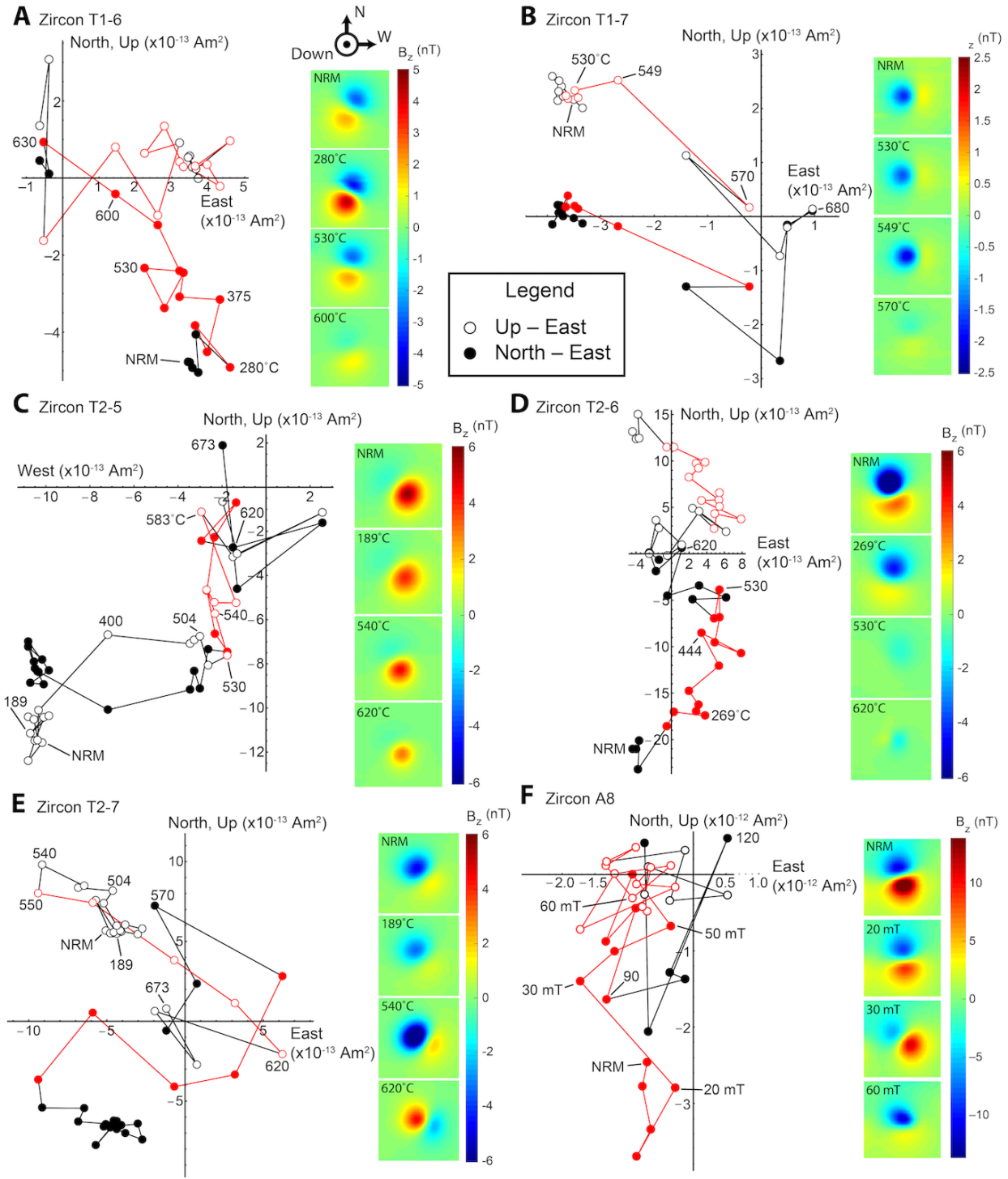

**Figure 2:** Orthogonal projection diagrams of single zircon demagnetization sequences. Single zircon thermal demagnetization sequences (**A-E**) show one or two component magnetizations. Panel C shows an example of a high noise demagnetization sequence that was not included in the final paleointensity analysis while panels A, B, D, and E show zircons with higher stability (see Table 1). Typical AF demagnetization sequence (**F**) shows significantly higher noise compared to thermal demagnetization. Solid and open circles denote the projection of magnetization in the horizontal and vertical planes, respectively. Red data points indicate range used in paleointensity calculations. Insets in each panel show the SQUID microscope maps of zircons at the indicated demagnetization steps [shown is the vertical component of the magnetic field at a height of ~230 μm, with positive (negative) field values representing out-of-plane (into-plane) field directions]. Paleointensity (Arai) diagrams for the same six zircons shown here are plotted in Fig. 4.

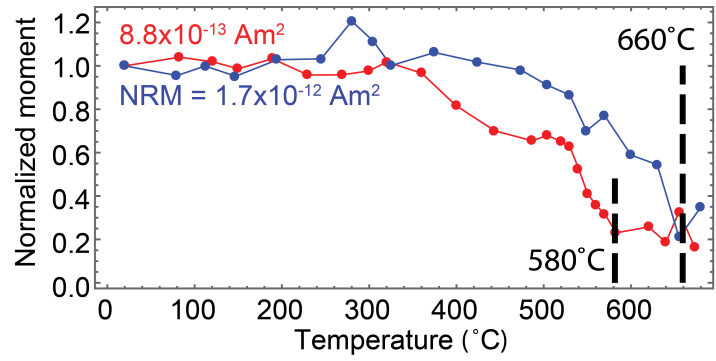

**Figure 3:** Thermal demagnetization of NRM for two zircons. The dominant ferromagnetic phase in samples T1-8 (blue) and T2-5 (red) are hematite and magnetite, respectively.

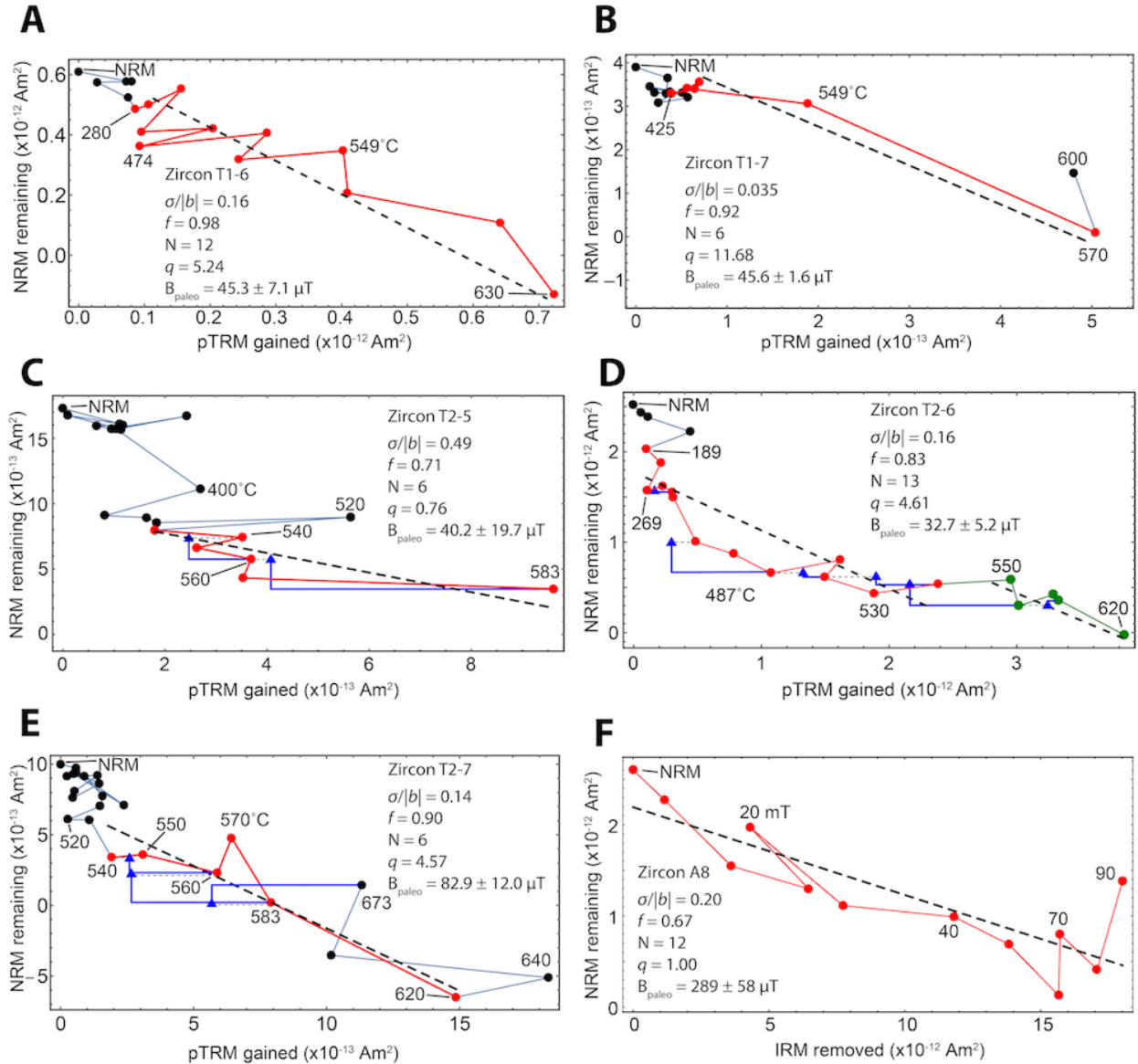

**Figure 4:** Paleointensity (Arai) diagrams for zircons subjected to (**A-E**) thermal- and (**F**) AF-based paleointensity analysis. Directional demagnetization data (Zijderveld diagrams) for these four zircons are shown in Fig. 2. Of the thermally analyzed zircons, T1-6 and T1-7 (panels A and B) were treated with sequential thermal demagnetization and pTRM acquisition while T2-5 through T2-7 (panels C-E) were measured using the IZZI Thellier-Thellier protocol including pTRM checks (blue triangles). Panel C shows data typical of zircons not included in the final

paleointensity mean due to high scatter. Dashed black lines show the least-squares fit to the chosen data range, indicated in red.

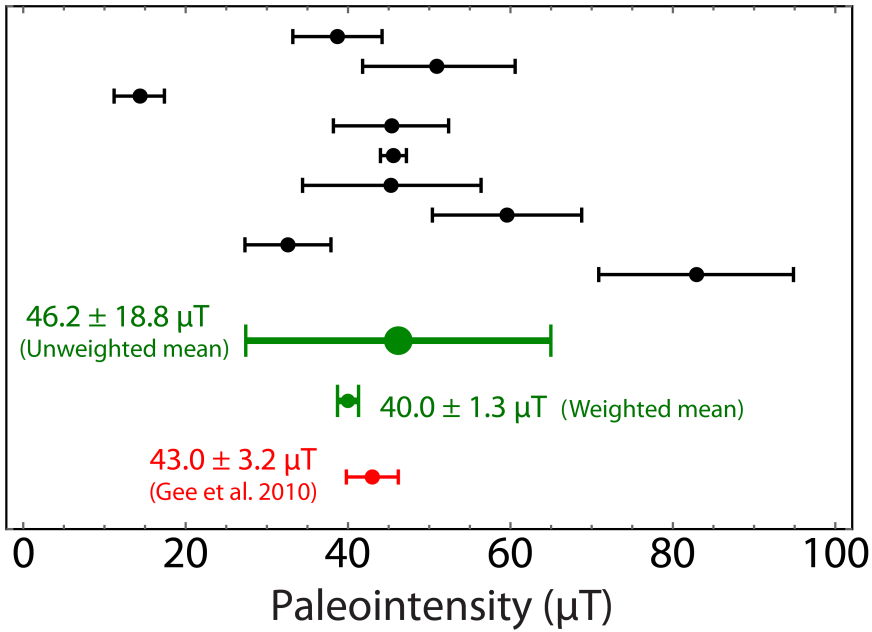

**Figure 5:** Paleointensity and 1σ uncertainty for all nine zircons included in the mean (black). Order from top to bottom corresponds to that shown in Tables 1 and 2. Unweighted and weighted means and 1σ uncertainties are shown in green while the mean paleointensity of bulk samples from Gee et al. (2010) is shown in red.

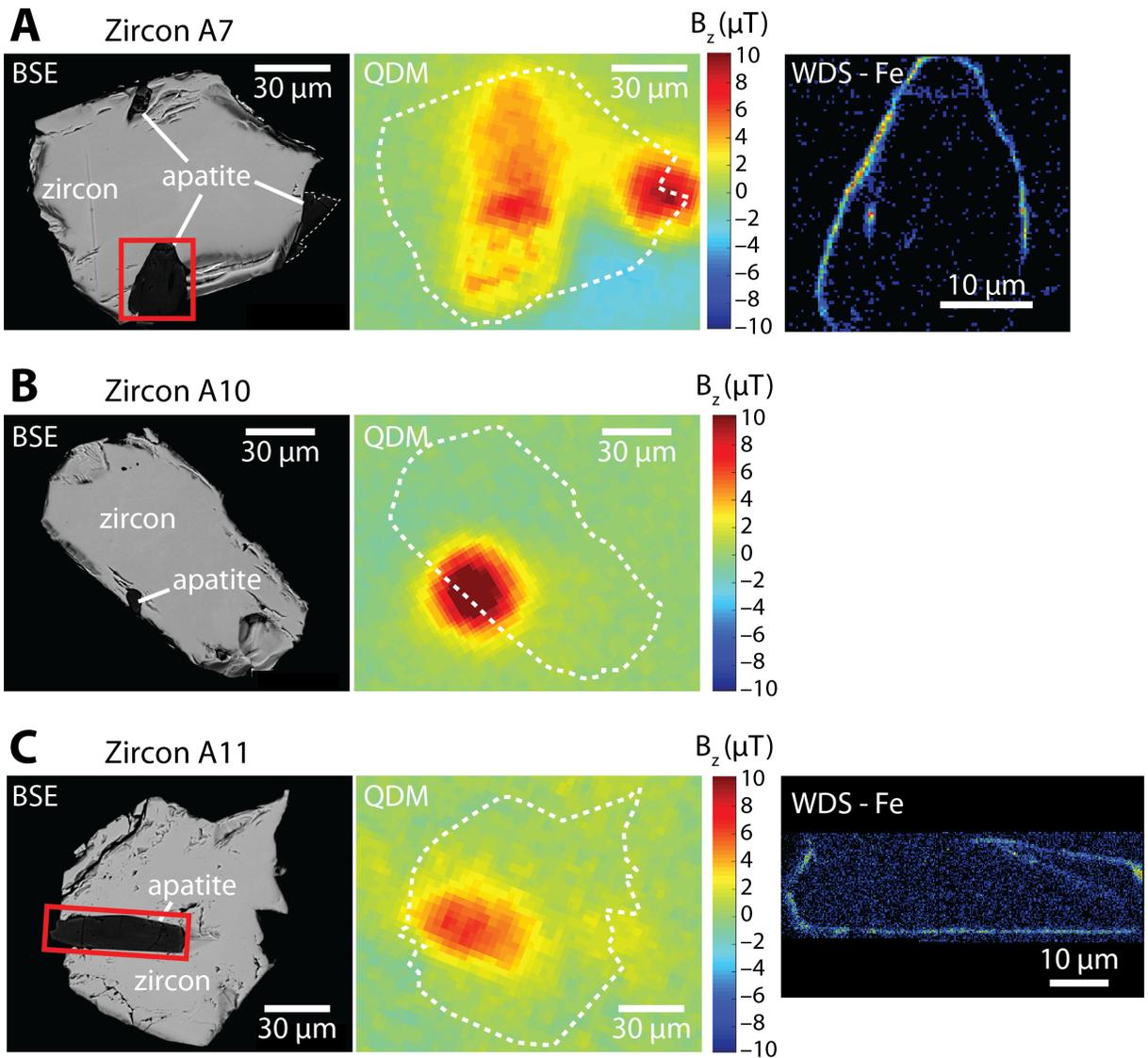

**Figure 6:** Backscatter election (BSE), quantum diamond microscope (QDM), and wavelength dispersion spectroscopy Fe element (WDS-Fe) maps of zircons showing the association of ferromagnetic sources with apatite inclusions. QDM maps were taken after a near-saturation IRM of 0.4 T was imparted in the out-of-plane (positive *z*) direction. Fields of view of WDS-Fe maps in panels A and C correspond to the red boxes in the BSE maps. Lighter colors reflect higher abundance of Fe in the WDS-Fe maps.

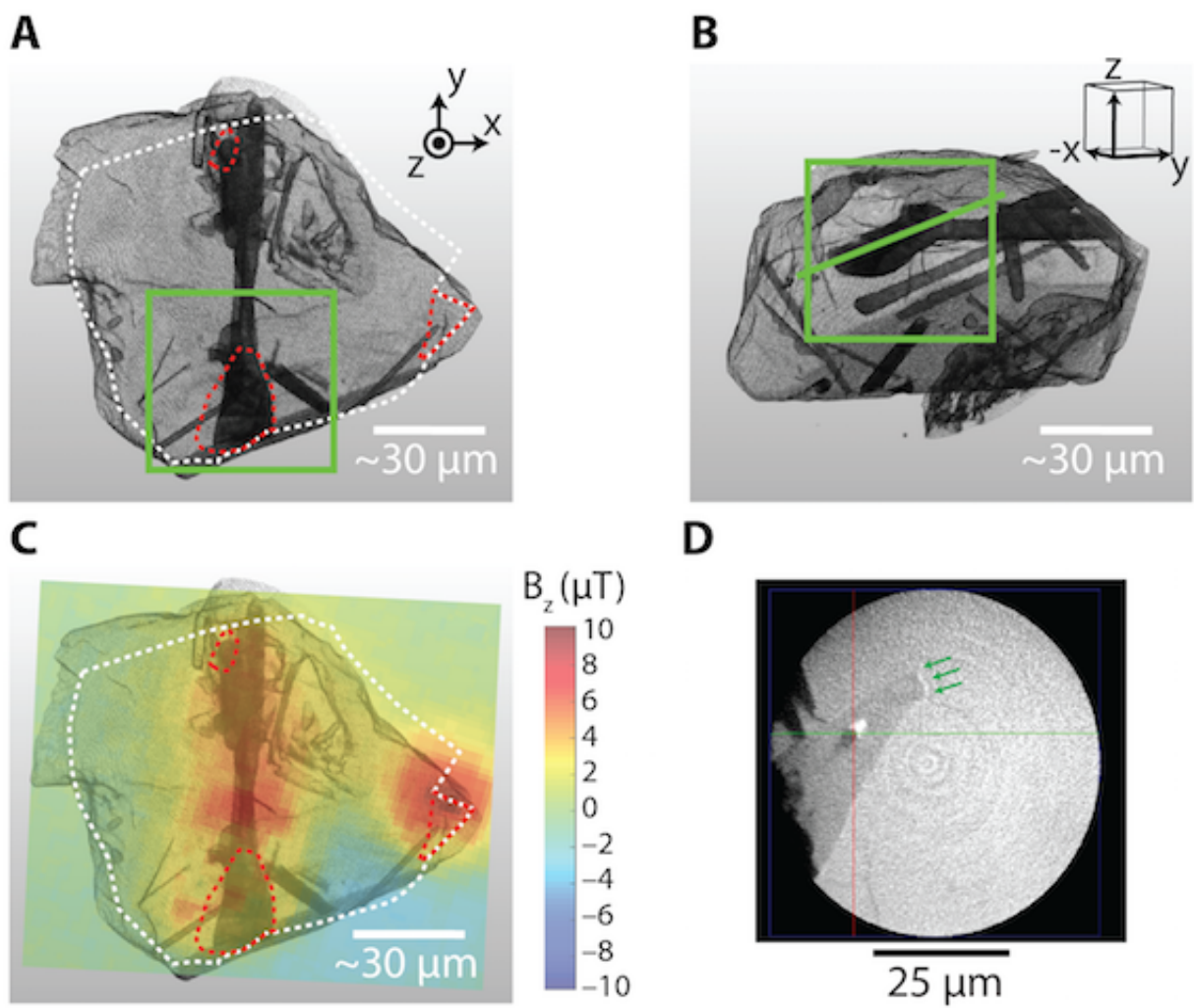

**Figure 7:** High-resolution X-ray tomography. (**A**, **B**) Three-dimensional renderings of zircon A7 in two different perspectives and (**C**) comparison to magnetic field map from QDM. Dashed white line in panels (A) and (C) approximate the location of the BSE-imaged surface in Fig. 6, while dashed red lines highlight the locations of surface apatite as identified from BSE and WDS analysis. Comparison to BSE and WDS maps indicate that light gray in 3D renderings represents zircon while dark gray represents apatite inclusions. Because the BSE maps are limited to the exposed surface, differences in perspective, and loss of material during remounting of zircon for X-ray tomography (including detachment of the rightmost apatite inclusion), the outline of the zircon from the BSE map does not match exactly that from the X-ray tomographic

renderings. Note the correspondence between the distribution of strong magnetization and internal, columnar apatite inclusions in panel (C). An animation of the 3D rendering is available in the Online Supplement. (**D**) A slice taken from high resolution X-ray tomography scan using the Zeiss Xradia Ultra 800. The field of view in panel D is highlighted by green boxes in panels A and B, with the image taken parallel to the green diagonal in B. Spatially correlated with the buried magnetic feature in the QDM image is a 2 μm Fe-oxide particle (bright white feature) appearing as an inclusion in the apatite (dark gray). Possible smaller Fe-oxide particles are highlighted with green arrows in panel D.

**Table 1:** Results of thermal demagnetization experiments on single zircons

| Sample name | Initial NRM ($\times 10^{-12}$ A m$^2$) | Main component range (°C) | MAD (°) | DANG (°) | Additional component range (°C) | MAD (°) | DANG (°) |
|---|---|---|---|---|---|---|---|
| T1-1 | 2.92 | 504-570 | 9.0 | 55.8 | 600-655 | 19.1 | 17.0 |
| T1-2 | 0.93 | 375-530 | 15.2 | 21.1 | | | |
| T1-3 | 5.72 | 20-530 | 7.2 | 7.5 | | | |
| T1-4 | 2.29 | 375-570 | 18.7 | 10.0 | | | |
| T1-5 | 1.55 | 325-549 | 16.8 | 73.2 | 549-655 | 27.8 | 7.35 |
| T1-6 | 0.61 | 280-630 | 20.1 | 11.1 | | | |
| T1-7 | 0.39 | 425-570 | 10.4 | 19.4 | 630-680 | 8.7 | 17.0 |
| T1-8 | 0.88 | 504-600 | 15.8 | 57.2 | | | |
| T2-1 | 0.18 | 520-640 (-583) | 18.6 | 14.7 | | | |
| T2-2 | 0.46 | 570-620 | 13.6 | 29.4 | | | |
| T2-3 | 0.25 | None | N/A | N/A | | | |
| T2-4 | 0.61 | 20-444 (-300) | 20.7 | 48.5 | | | |
| T2-5 | 1.73 | 504-583 | 24.2 | 19.6 | 360-504 | 17.3 | 52.4 |
| T2-6 | 2.52 | 189-540 | 19.4 | 30.0 | 540-620 | 38.3 | 7.4 |
| T2-7 | 1.00 | 540-620 | 17.9 | 9.8 | 640-673 | 30.1 | 7.6 |
| T2-8 | 0.57 | 20-530 | 14.0 | 26.4 | 560-640 | 17.1 | 12.5 |
| T2-9 | 1.50 | 20-640 | 26.1 | 22.3 | | | |
| T2-10 | 0.05 | None | N/A | N/A | 20-150 | 19.7 | 6.89 |

Notes: Only the "main component" of magnetization observed in each zircon (column three) is interpreted as a primary TRM and used for paleointensity analysis. Temperature limits for each component are based on the identification of magnetization components from the demagnetization sequence. MAD in columns 4 and 7 refers to the maximum angular deviation, which quantifies the scatter of

demagnetization steps around the mean direction. DANG in columns 5 and 8 refers to the deviation angle, which is the angular separation between the best-fit direction of the magnetization component with and without requirement for the fitted line to pass through the origin.

**Table 2:** Results of dual-heating paleointensity experiments on single zircons

| Sample name | Normalized scatter ($\sigma/|b|$) | NRM fraction ($f$) | Gap factor ($g$) | Quality index ($q$) | Number of steps (N) | Paleointensity (µT) | $1\sigma$ uncertainty (µT) | Included in mean |
|---|---|---|---|---|---|---|---|---|
| T1-1 | 0.14 | 0.61 | 0.52 | 2.26 | 4 | 38.7 | 5.5 | Yes |
| T1-2 | 0.34 | 0.52 | 0.72 | 1.08 | 5 | 50.9 | 17.8 | No |
| T1-3 | 0.18 | 1.19 | 0.87 | 5.65 | 13 | 51.2 | 9.4 | Yes |
| T1-4 | 0.27 | 1.03 | 0.57 | 2.16 | 7 | 43.6 | 11.8 | No |
| T1-5 | 0.22 | 0.47 | 0.80 | 1.72 | 6 | 14.2 | 3.1 | Yes |
| T1-6 | 0.16 | 0.98 | 0.84 | 5.24 | 12 | 45.3 | 7.1 | Yes |
| T1-7 | 0.035 | 0.92 | 0.43 | 11.68 | 6 | 45.6 | 1.6 | Yes |
| T1-8 | 0.24 | 0.80 | 0.72 | 2.37 | 5 | 45.4 | 11.0 | Yes |
| T2-1 | 0.15 | 0.57 | 0.36 | 1.34 | 8 | 59.6 | 9.2 | Yes |
| T2-2 | 0.45 | 0.38 | 0.42 | 0.36 | 3 | 59.1 | 26.7 | No |
| T2-3 | N/A | N/A | N/A | N/A | N/A | N/A | N/A | No |
| T2-4 | 0.27 | 0.42 | 0.72 | 0.80 | 11 | 88.6 | 24.2 | No |
| T2-5 | 0.63 | 0.71 | 0.68 | 0.76 | 8 | 34.9 | 22.0 | No |
| T2-6 | 0.16 | 0.83 | 0.88 | 4.61 | 13 | 32.7 | 5.2 | Yes |
| T2-7 | 0.14 | 0.90 | 0.74 | 4.57 | 6 | 82.9 | 12.0 | Yes |
| T2-8 | 0.58 | 0.61 | 0.56 | 0.59 | 16 | 67.3 | 39.3 | No |
| T2-9 | 0.12 | 0.73 | 0.63 | 13.88 | 23 | 126.8 | 4.2 | No |
| T2-10 | N/A | N/A | N/A | N/A | N/A | N/A | N/A | No |
| Mean | | | | | | 46.2 | 18.8 | |

Notes: Only analysis of the main magnetization component of each zircon is shown here. Inclusion in overall mean required $\sigma/|b| \leq 0.25$. Sample T2-9 was also excluded due to possible strong field contamination. See text for details. Column 2 gives the $1\sigma$

uncertainty of the best-fit slope in the Arai diagram normalized by the slope value. Column 3 gives the magnitude of the NRM over which the paleointensity fit is performed as a fraction of the total as defined by Coe et al., (1978). Column 4 and 5 quantify the evenness of spacing between fitted points in the Arai diagram and the overall quality factor, respectively, as defined by Coe et al., (1978). Column 6 gives the number of demagnetization step included in the paleointensity analysis.

**Table 3:** Summary of pTRM check results

| Sample | $\Delta pTRM_{max}$ ($10^{-13}$ Am$^2$) | $\Delta pTRM_{max}/\sigma_{pTRM}$ | $DRAT_{max}$ | Fit range $\Delta pTRM_{max}$ ($10^{-13}$ Am$^2$) | Fit range $\Delta pTRM_{max}/\sigma_{pTRM}$ | Fit range $DRAT_{max}$ |
|---|---|---|---|---|---|---|
| T2-1 | 0.79 | 2.21 | 1.08 | 0.31 | 0.87 | 0.14 |
| T2-2 | 3.41 | 1.58 | 0.47 | 2.18 | 0.43 | 0.30 |
| T2-3 | 1.78 | 1.61 | N/A | N/A | N/A | N/A |
| T2-4 | 2.52 | 0.70 | 1.07 | 0.59 | 0.16 | 0.25 |
| T2-5 | 4.08 | 1.92 | 0.43 | 4.08 | 1.92 | 0.43 |
| T2-6 | 4.23 | 0.49 | 0.26 | 1.85 | 0.22 | 0.11 |
| T2-7 | 3.23 | 1.57 | 0.13 | 3.23 | 1.57 | 0.13 |
| T2-8 | 1.02 | 0.77 | 0.28 | 1.02 | 0.77 | 0.28 |
| T2-9 | 3.75 | 4.80 | 0.31 | 0.63 | 0.80 | 0.05 |
| T2-10 | 2.94 | 4.10 | N/A | N/A | N/A | N/A |

Notes: In column 2, the quantity $\Delta pTRM_{max}$ represents the largest difference between repeated pTRM acquisitions at the same temperature for the sample. Column 3 gives the normalized values of $\Delta pTRM_{max}$ divided by $\sigma_{pTRM}$ (the standard deviation of difference between pTRM gained and the best-fit value from the Arai diagram, see text). Column 4 gives the highest absolute value of the DRAT parameter as defined by Selkin & Tauxe (2000). The final three columns show values of each parameter within the temperature range used to fit for paleointensity. For a pTRM check to be included, highest demagnetization step achieved before the pTRM check must fall within the fit range. Zircons T2-3 and T2-10 had no identifiable components on which to compute paleointensity.

**Supplementary Data:**

Evaluating the paleomagnetic potential of single zircon crystals using the Bishop Tuff

Roger R. Fu, Benjamin P. Weiss, Eduardo A. Lima, Pauli Kehayias, Jefferson F. D. F. Araujo, David Glenn, Jeff Gelb, Joshua F. Einsle, Ann M. Bauer, Richard J. Harrison, Guleed A.H. Ali, Ronald L. Walsworth

**Discussion of methodology for constraining the bulk zircon magnetization**

As described in the section entitled "*Microscopy of Bishop Tuff zircons*," we use high-resolution QDM maps of the magnetic field from a zircon (A15) magnetized in a near-saturation 0.4 T magnetic field to constrain the possible moment contribution from volumetrically dispersed ferromagnetic sources in the zircon [e.g., Timms et al., (2012)]. Here we discuss the uncertainties of this analysis and argue that it is sufficient to produce a reliable, conservative upper bound on the intensity of bulk magnetization. The minimal amount of polishing conducted for the epoxy-mounted zircons resulted in the removal of only a small fraction of the total volume, resulting in sub-equant zircons that have aspect ratios rarely greater than 2:1 (Fig. 7). Even so, modeling the zircon as a uniformly magnetized sphere is an approximation. Furthermore, the subtraction of the fitted localized sources assume that these are well-modeled by a dipole; therefore, any non-dipolarity of the actual localized sources would result in the incomplete or excessive subtraction of the signals. A more comprehensive analysis would study the magnetic field from a zircon with no localized sources, avoiding any uncertainties introduced by the subtraction of localized sources. Furthermore, the QDM image in Fig. 6 has a field gradient from the applied magnetic field required by the QDM measurement with a full range of 0.46 µT across the image. Subtracting this gradient leaves a uniform background field and yields a weaker value for the bulk zircon dipole moment. However, such a subtraction may also remove some field contribution from the bulk magnetization, leading to an overly stringent limit. We chose the more conservative approach and refrain from subtracting the background gradient, strengthening the assertion that ~1% is an upper bound for the contribution of the bulk magnetization to the total zircon moment. Future analyses can more accurately constrain the bulk magnetization by modeling realistic zircon geometries, studying zircons with no localized sources near the surface, improving the QDM applied magnetic field gradient, and distinguishing localized and bulk sources by successively polishing or raising the diamond sensor height in 10 µm steps.

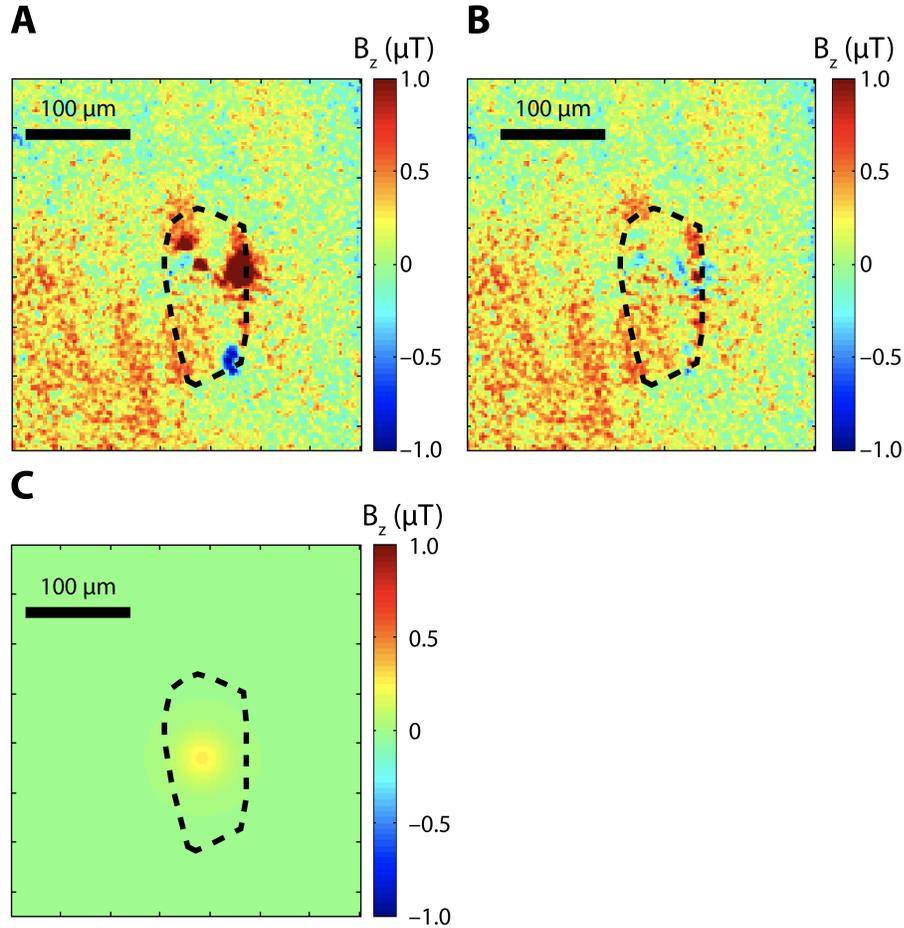

**Figure S1:** Elimination of localized ferromagnetic source signals and inversion for the maximum bulk magnetization moment of zircon A15, which has been given a 0.4 T near-saturation IRM in the out-of-plane direction. (**A**) Raw QDM image of the out-of-plane magnetic field component where the dashed black line denotes the outline of the zircons. (**B**) Processed QDM image where the four strongest localized magnetic moment sources have been fitted assuming a dipole source and removed. (**C**) Magnetic field corresponding to the best-fit dipole moment source for the map in panel (B) assuming source location at the center of the zircon and 40 μm depth. This represents an approximate upper bound to the magnetic moment due to volumetrically dispersed ferromagnetic grains.

**Table S1:** Results of AF demagnetization of single zircons and IRM paleointensity experiments

| Sample name | MAD (°) | Normalized scatter ($\sigma/|b|$) | NRM/IRM | Paleointensity (μT) | 1σ uncertainty (μT) |
|---|---|---|---|---|---|
| A1 | 29.1 | 0.18 | 0.28 | 826 | 146 |
| A2 | 26.1 | 0.32 | 0.03 | 88.4 | 28 |
| A3 | 22.1 | 0.21 | 0.16 | 465 | 99 |
| A4 | 13.7 | 0.12 | 0.21 | 620 | 76 |
| A5 | 44.8 | 0.22 | 0.08 | 236 | 51 |
| A6 | 15.2 | 0.08 | 0.06 | 192 | 14 |
| A7 | 26.1 | 0.31 | 0.11 | 335 | 103 |
| A8 | 22.3 | 0.20 | 0.10 | 289 | 58 |
| A9 | 20.5 | 0.15 | 0.31 | 915 | 135 |
| A10 | 30.2 | 0.15 | 0.27 | 819 | 126 |
| A11 | 30.4 | N/A | N/A | N/A | N/A |
| A12 | 33.2 | 2.57 | 0.01 | 22.4 | 57.7 |
| Mean | | | 0.15 | 437 | 314 |

Notes: Due to zircon A11's close location to the much stronger zircon A12 on the sample mount, we were unable to recover a reliable IRM demagnetization sequence. For the same reason, we were unable to recover reliable NRM or IRM data from three further zircons (A13-A15), which are not listed here. AF range fitted for the NRM to near-saturation IRM ratio (NRM/IRM) is 0 to 90 mT for all samples. The number of fitted steps is 12 for all samples. IRM paleointensities are computed using the slope of the least squares best fit line and an empirical factor of 3000 μT (Gattacceca and Rochette, 2004; Kletetschka et al., 2003). Column 2 gives the maximum angular deviation, which describes the scatter of demagnetization data around the best-fit direction. Column 3 gives the 1σ uncertainty of the best-fit slope in the Arai diagram normalized by the slope value.